# Colloidal quantum dot infrared lasers featuring sub-single-exciton threshold and very high gain

Nima Taghipour[1], Mariona Dalmases[1], Guy L. Whitworth[1], Andreas Othonos[2], Sotirios Christodoulou[3], and Gerasimos Konstantatos[1,4]*

[1]ICFO, Institut de Ciències Fotòniques, The Barcelona Institute of Science and Technology, 08860, Castelldefels, Barcelona, Spain

[2]Laboratory of Ultrafast Science, Department of Physics, University of Cyprus, Nicosia 1678, Cyprus

[3]Inorganic Nanocrystals Laboratory, Department of Chemistry, University of Cyprus, 1678 Nicosia, Cyprus

[4]ICREA, Institució Catalana de Recerca i Estudis Avançats, 08010 Barcelona, Spain

*Gerasimos.konstantatos@icfo.eu

**The use of colloidal quantum dots (CQDs) as a gain medium in infrared laser devices has been underpinned by the need for high pumping intensities, very short gain lifetimes and low gain coefficients. Here, we employ PbS/PbSSe core/alloyed-shell CQDs as an infrared gain medium that results in highly suppressed Auger recombination with a lifetime of 485 ps, lowering the amplified spontaneous emission (ASE) threshold down to 300 µJ.cm$^{-2}$, and showing a record high net modal gain coefficient of 2180 cm$^{-1}$. By doping these engineered core/shell CQDs up to nearly filling the first excited state, we demonstrate a significant reduction of optical gain threshold, measured by transient absorption, to an average–exciton population per dot $\langle N^{th} \rangle_g$ of 0.45 due to bleaching of the ground state absorption. This in turn have led us to attain a 5-fold reduction in ASE threshold at $\langle N^{th} \rangle_{ASE}$= 0.70 excitons per dot, associated with a gain lifetime of 280 ps. Finally, we use these heterostructured QDs to achieve near-infrared lasing at 1670 nm at a pump fluences corresponding to sub-single-exciton per dot threshold ($\langle N^{th} \rangle_{Las}$ = 0.87). This work brings infrared CQD lasing thresholds on par to their visible counterparts, albeit starting with an 8-fold degenerate system and paves the way towards solution processed infrared laser diodes.**

Colloidal quantum dots (CQDs) have been considered as versatile and promising gain media for solution-processed lasers with tunable size-dependent emission[1–11]. Optical gain in these CQDs is realized when the number of carriers in the excited state is greater than the number of carriers in the ground state, a condition known as population inversion. Therefore, a fundamental challenge in the advancement of technologically viable CQD-laser devices is the multiexcitonic nature of light amplification[1,4,7,10], resulting in high lasing thresholds. This challenge is especially problematic for infrared-emitting PbS(Se) CQDs which have an 8-fold degeneracy at the band-edge, meaning that population inversion is satisfied when the ensembled-average number of excitons occupancy per dot, ⟨N⟩, is greater than approximately four[1,2,10,12]. The required high number of excitons within a CQD leads to an unfavorable, competitive-to-lasing process known as Auger recombination, where the exciton energy is transferred to a third carrier that subsequently undergoes non-radiative relaxation[7,10,13], leading to optical gain lifetime, $\tau_g$, of tens of picoseconds in standard CQDs[1,7,12,13].

Room-temperature tunable infrared amplified spontaneous emission (ASE) and lasing have been recently demonstrated in PbS-[1,2,12,14], and $Ag_2Se$-based CQDs[15]. Particularly, PbS-based CQDs show ASE/lasing thresholds in the range of 500-2000 µJ.cm$^{-2}$,[1,2,14] a gain lifetime of 40 ps[1], and a limited gain coefficient of 120 cm$^{-1}$.[1,2] As previously stated, these high lasing thresholds and very short optical gain lifetimes stem from fast Auger recombination in view of the high degeneracy of PbS CQDs. Previously, in visible-emitting nanocrystals, it has been experimentally proven that Auger processes can be suppressed by using a core/shell heterostructure, thereby reducing the lasing threshold[16–19]. Another approach for reducing the lasing threshold is by populating the conduction band of the CQDs with permanent charges, thus partially/totally bleaching the ground state absorption[1,13,20]. This method has also been shown to prolong the gain lifetime since the Auger process of a charged exciton is reduced compared to that of multiexcitons on a neutral CQD [11,20]. Here, we demonstrate sub-single-exciton infrared lasing with record low threshold by using heavily doped core/shell infrared CQDs as the gain medium. In doing so, we prepared PbS/PbSSe core/alloyed-shell (C/A-S) CQD films, in which suppressed Auger decay is combined with highly efficient doping. Using C/A-S CQDs allowed us to nearly fill the first excited state with more than 7 permanent electrons on average per dot. In our C/A-S CQDs, the biexciton Auger lifetime is prolonged by 2.5 times compared to core CQDs. Consequently, optical gain lifetime in doped devices is at least 7-times longer than any previously reported infrared optical gain lifetime in CQD systems. As a result, we demonstrate more than one order of magnitude reduction in the ASE threshold, where the gain coefficient in our devices exceeds (up to 20-fold) those reported any solution-processed infrared gain media at room temperature so far.

In this work, we synthesized PbS/PbSSe core/shell CQDs by the hot-injection method, with PbS core size of ~ 4.85 nm (see Methods) and a subsequent thin alloyed shell (~1.10 nm) of PbSSe grown around. This alloy shell effectively smooths the quantum potential confinement which has been previously shown to reduce the strength of the intraband transitions involved in non-radiative Auger recombination[21]. We illustrate the schematic of the core/shell structure in Fig. 1a, while the absorption spectra of core and core/shell CQDs are shown in Fig. 1b. Because of the relaxation of the quantum confinement, the excitonic peak of the CQDs is shifted to lower energies by growing the outer shell[11,16,17,19]. Transmission electron microscopy (TEM) images (Fig. 1c) show well-defined crystalline particles and STEM-EDS elemental mapping images (Fig. 1d) corroborate the formation of the shell. X-ray powder diffraction (XRD) characterizations of the core/shell CQDs provided in Supplementary Fig. S3. In order to permanently dope our solid films, we carried out a ligand exchange process, using 1-ethyl-3-methylimidazolium iodide (EMII) as a ligand, in this process the $S^{2-}$ were substituted by $I^-$ on the (100) crystalline facet of the CQDs, as represented schematically in Supplementary Fig. S5. Subsequently, a thin layer of alumina is deposited via atomic layer deposition (ALD) to protect the treated surface of CQDs from atmospheric electron acceptors (oxygen/water)[1](see Methods). This iodide ligand exchange procedure populates the conduction band of the CQDs with permanent electrons. Bleaching of the first excitonic feature due to Pauli blocking is a clear signature of doping the CQD films (Supplementary Fig. S6). From here on $\langle n_e \rangle$, will be defined as the average number of doping electrons per quantum dot in the films, calculated from the quenching of absorption in the area of first excitonic transition (see Supplementary Note 2).

**Observation of sub-single-exciton gain in doped C/A-S CQDs**

To evaluate the optical gain performance of core and C/A-S CQDs, we carried out ultrafast transient absorption (TA) spectroscopy on thin films (see Methods). To quantify the optical gain threshold, we plot $|\Delta\alpha/\alpha_o|$ as a function of $\langle N \rangle$ at ASE wavelengths in Fig. 2a-c. The optical gain threshold in terms of average number of excitons per dot, $\langle N^{th} \rangle_g$, is determined from the intersection of the experimental data (symbol) with horizontal line shown by $|\Delta\alpha/\alpha_o| = 1$. For the neutral PbS core CQDs sample, this was measured to be $\langle N^{th} \rangle_g$ = 4.40 (Fig. 2a), which is close to theoretical value of 4.04 expected from Poisson statistics (see Supplementary Fig. S4). The details of calculating average number of excitons described in Supplementary Note.3. The gain threshold in neutral PbS/PbSSe C/A-S CQD sample drops to 3.80, ascribed to the suppression of Auger recombination and the reduction in self-absorption losses. As previously shown, a large shift of the stimulated emission spectrum with respect to the excitonic absorption reduces the re-absorption losses in the materials[4,11,17,19]. The TA characterization on heavily doped C/A-S CQDs (with $\langle n_e \rangle \approx 7.40$) shows a significant reduction of the gain threshold to $\langle N^{th} \rangle_g \approx 0.45$. This observation demonstrates a considerable, almost 10-fold, reduction of the optical gain threshold in heavily doped C/A-S CQDs compared to the neutral core CQDs, which paves the way for sub-single-exciton infrared lasing from an 8-fold degenerate PbS-based CQD system. Another important observation in using PbS/PbSSe C/A-S CQDs as a gain medium is the prolongation of the biexciton Auger lifetime (Fig. 2d), measured to be is ~485 ps in C/A-S CQDs, while core CQDs shows a value of ~230 ps. In Fig. 2e, we plot the sum of the linear absorption, $\alpha_0$, and the TA dynamics, $\Delta\alpha(t)$ at the corresponding ASE wavelengths, to measure the time for which gain persists in the CQDs ($\alpha_0 + \Delta\alpha < 0$), denoted as the gain lifetime $\tau_g$. For the heavily doped C/A-S CQDs ($\langle n_e \rangle \approx 7.40$) was measured $\tau_g$ to be ~280 ps which is remarkably longer than that of the neutral core CQDs (40 ps) and the neutral C/A-S CQDs (60 ps). Thus, the synergism of C/A-S CQDs with very strong doping successfully addresses the issues of high optical gain thresholds and very fast optical gain relaxation in infrared-emitting CQDs.

**ASE and gain coefficient characterizations**

Next, we examined the optical gain performance of undoped and doped C/A-S CQDs by conducting ASE measurements on thin films. To do so, we excited the samples with fs-laser using stipe excitation and the emission was collected perpendicular to the excitation axis as explained in the Method section (see Supplementary Fig. S7). The collected emission as a function of the excitation fluence shows a clear transition from spontaneous emission to ASE through an abrupt change in the slope of the output intensity and the spectral narrowing (Fig. 3a, b). We display the output intensity as a function of pump fluence in Fig. 3a, in which the ASE thresholds for undoped-C/A-S CQDs film is measured as ~300 μJ.cm$^{-2}$, while the ASE threshold for the similar size undoped-core sample is ~500 μJ.cm$^{-2}$. Interestingly, by doping the C/A-S film ($\langle n_e \rangle$ = 7), the ASE threshold drops by 70% to ~90 μJ.cm$^{-2}$, which represents an improvement of almost one order of magnitude from commonly reported values (500-800 μJ.cm$^{-2}$) for similar emission wavelengths[1,2,12]. The measured ASE thresholds in terms of average number of excitons per dot, $\langle N^{th} \rangle_{ASE}$, correspond to a value of 3.80 (before doping) and 0.70 (after doping). The observation of ASE at sub-single-exciton-per-dot population regime for the doped system suggests that the gain originates from charged single-excitons, for which the estimated gain threshold is 0.56 (see Supplementary Note 1). It is worth mentioning that the attained ASE thresholds in our C/A-S CQDs is more than one order of magnitude lower than corresponding core QDs (> 1 mJ.cm$^{-2}$)[2]. In general, the observed lower ASE thresholds in C/A-S compared to core-only CQDs is due to increasing absorption cross-section, suppressing of Auger recombination (see Fig. 2d) and reducing the re-absorption losses. Particularly, spectral analysis of the ASE spectra (Fig. 3b) suggests that the ASE peak of C/A-S CQDs is 32 meV shifted to longer wavelengths with respect to the first excitonic transition, while this value is 18 meV (see Supplementary Fig. S8) for core-only CQDs with similar ASE wavelengths. Lowering the ASE threshold owing to reducing the self-absorption has been reported previously in visible emitting -CQDs[4], -colloidal quantum wells[17] and organic-dye[22] gain media.

Another important benchmarking parameter for the laser devices is the net modal gain coefficient ($g_{net}$) of the gain medium, which quantifies light amplification per unit length and is of paramount importance for integrated and compact laser diodes. Recent advances in visible-emitting CQDs led to achieve $g_{net}$ values as high as 6600 cm$^{-1}$ for colloidal nanoplatelets, and 2800 cm$^{-1}$ for core/shell CQDs. However, to date infrared counterparts have reached much lower $g_{net}$ of less than 120 cm$^{-1}$ for PbS core CQDs,[1,2] and 2.4 cm$^{-1}$ for HgTe CQDs.[3] We thus sought to measure the net modal gain coefficient of our C/A-S CQDs, characterized by the variable stripe length (VSL) method. The sample was excited by a stripe-shaped beam (Supplementary Fig. S7), where the length of stripe was controlled by an adjustable slit (see Methods). As an exemplary case, we display the integrated emission intensity of the ASE peak in Fig. 3c against stripe length for the case doped C/A-S CQDs thin film. The integrated emission intensity raises exponentially as a function of the stripe length. This relation is generally expressed as $I(l) = A(e^{g_{net}l} - 1)$,[23] in which $l$ is the stripe length, and $g_{net}$ is the net modal gain coefficient. Using this equation, we extract $g_{net}$ as 2180± 180 cm$^{-1}$ at a fluence of 2200 µJ.cm$^{-2}$. VSL measurements at different pump fluences are provided in Supplementary Fig. S9. Fig. 3d shows that the net modal gain increases with pump fluences in the range of 200-750 µJ.cm$^{-2}$, then deviates from this behavior and saturates at almost 2000 cm$^{-1}$ for higher pump fluences. This behavior is consistent with ASE measurements (Fig. 3a), in which the ASE signal output also saturates at high fluences. Our achieved $g_{net}$ sets a record among any other solution-processed infrared gain material, and is found to be on par with epitaxially grown infrared semiconductors such as GaInAs/InP quantum dots.[24] This high gain coefficient in PbS/PbSSe C/A-S (thin shell) CQDs can be ascribed to three factors: i) PbSSe shell effectively suppresses the nonradiative Auger processes, ii) A large-core/thin-shell architecture maximizes the volume fraction of the emitting PbS core over the PbSSe shell, and iii) the attractive Coulombic exciton–exciton interactions shift ASE away from the ground-state absorption (Fig. 3b)[19].

**Effect of CQDs doping on lasing threshold**

Lastly, following the successful realization of sub-single-exciton stimulated emission using doped-C/A-S CQDs, we aimed to fabricate the first solution processed infrared laser with sub-single-exciton threshold. To do so, we deposited the CQDs on top of nanostructured gratings in order to make second-order distributed feedback (DFB) structures. The DFB grating parameters were precisely designed to provide in-plane resonance at the ASE peak, the fabrication of which and lasing characterizations are given in Methods. We depict the integrated spectral output of lasing emission as a function of the excitation fluence in Fig. 4a for undoped- and doped-CQD/DFB devices, indicating the lasing thresholds of ~430 and 160 µJ.cm$^{-2}$, respectively. To quantify the lasing thresholds in terms of $\langle N \rangle$ (see Supplementary Note. 3), we assumed a packing density of 74% for the CQDs thin film, and with a measured 19% absorption at the excitation wavelength 1030 nm, the latter fluence equals to an average exciton occupancy per CQD of $\langle N^{th} \rangle_{Las}$= 0.87. Fig. 4b shows the collected lasing spectra (perpendicular to the surface) above the lasing threshold for neutral- (left panel) and doped-C/A-S CQDs (right panel) at $\langle N \rangle$= 5.07 and 1.05, respectively. The single-mode lasing of neutral device is centered at 1660 nm within a linewidth of ~ 0.9 nm (0.41 meV), while doped device operates at 1670 nm having a linewidth of ~ 1 nm (0.46 meV). Optical images of the surface of the CQD-DFB device below and above the threshold given in Fig. 4c. The laser device exhibits strong linear polarization with a ratio of polarization of $(I_\parallel - I_\perp)/(I_\parallel + I_\perp)$= 0.94, where $I_\parallel$, $I_\perp$ are the lasing intensity parallel and perpendicular to the optical axis, correspondingly (Fig. 4d). Finally, our laser device operated at room temperature for more than 8 hours under uninterrupted pumping at 10 kHz (Fig. 4e), setting a new stability record for a CQD laser. Over this time, we did not witness any thermal or optical damage to the laser device.

**Conclusion**

We have reported sub-single exciton gain and lasing in an infrared CQD laser, matching the performance of previously reported visible counterparts, albeit infrared CQDs possess an 8-fold state degeneracy. This was achieved by employing a PbS/PbSSe core-alloyed shell QD heterostructure doped up to almost filling the first excited state with 7 electrons per dot. This synergistic effect facilitated suppressed Auger recombination evidenced by a gain lifetime of ~300 ps and very high gain on the order of $10^3$ cm$^{-1}$. Our work facilitates a path towards practical infrared laser devices for use in LIDAR applications, photonic integrated circuits (PICs) and eye-safe optical telecommunications.

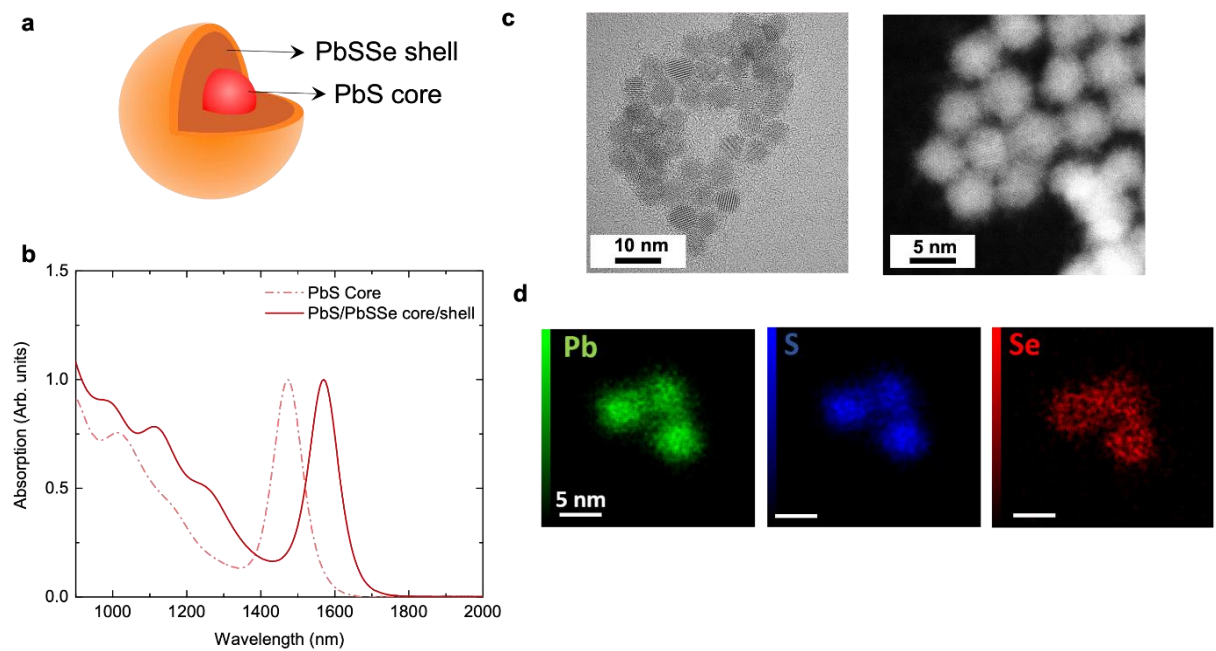

**Figure 1 | Structural characterization of PbS/PbSSe core/alloyed-shell (C/A-S) CQDs structure.**
**a,** Schematic representation of core/shell architecture. **b,** Absorption spectra of PbS core and PbS/PbSSe C/A-S CQDs. **c,** High resolution TEM (left) and High-angle annular dark field STEM (right) images of C/A-S CQDs. **d,** Energy dispersive X-ray spectroscopy (EDS) elemental maps of C/A-S CQDs. Additional TEM images provided in Supplementary Fig. S1&S2

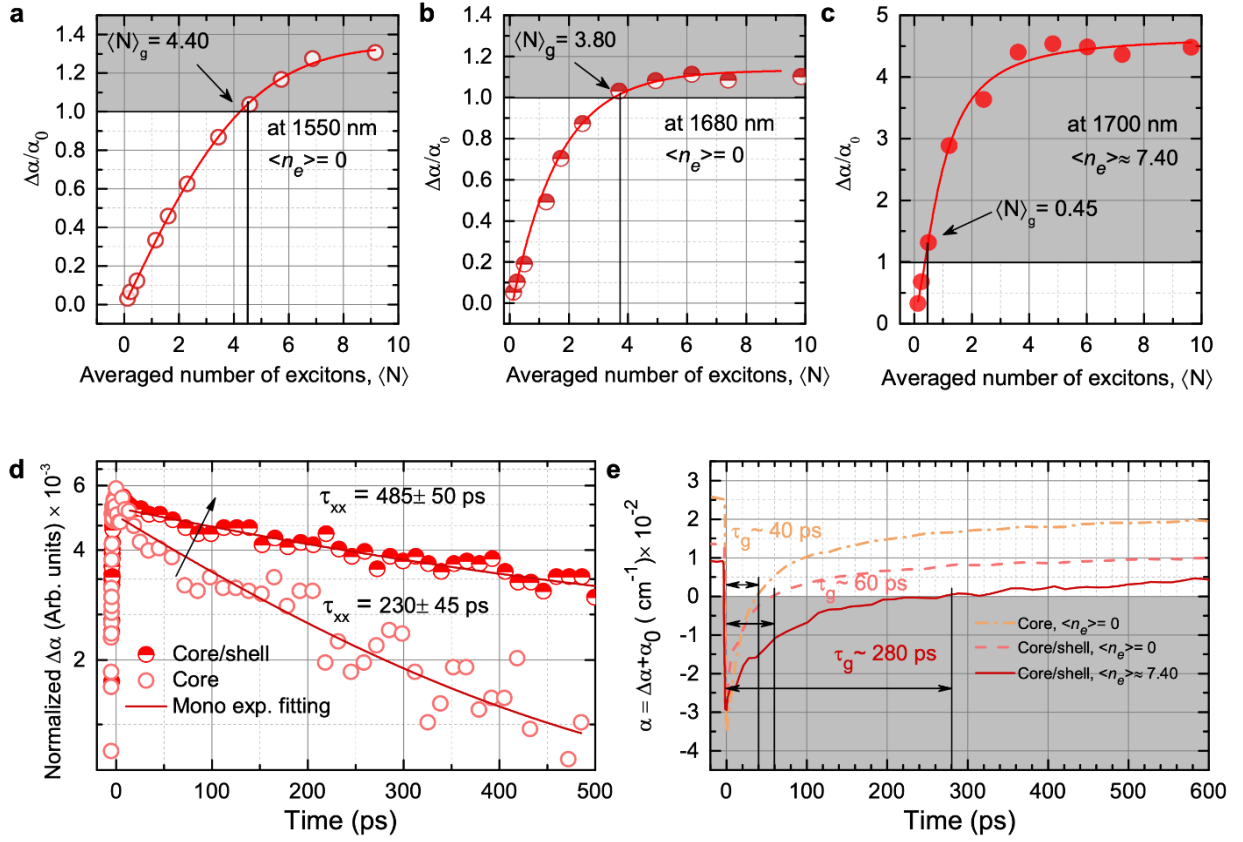

**Figure 2 | Optical gain in undoped and doped core and C/A-S CQDs.** Normalized absorption bleaching $|\Delta\alpha/\alpha_o|$ as a function of $\langle N \rangle$ ($\langle N \rangle$ is the average number of excitons per dot) at the corresponding ASE peak position for **(a)** undoped solid film of PbS core CQDs, **(b)** undoped and **(c)** doped solid film of PbS/PbSSe C/A-S CQDs. The solid lines are for the eye guidance. The shaded region corresponds to optical gain regimes $|\Delta\alpha/\alpha_o| > 1$ which implies a gain threshold of $\langle N_g \rangle$= 4.40, 3.80 and 0.45 for undoped core **(a)**, undoped C/A-S **(b)** and doped C/A-S CQDs **(c)**. **d,** Dynamics of bi-excitons (symbol), obtained from TA at the corresponding ASE peak. The red solid line shows single exponential fitting to experimental data. **e,** Nonlinear absorption ($\alpha = \alpha_o + \Delta\alpha$) as a function of time delay at the ASE peak position for undoped core, undoped C/A-S and doped C/A-S CQDs, indicating net optical gain lifetime of $\tau_g \approx$ 40, 60 and 280 ps, respectively. The shaded region shows optical gain ($\alpha < 0$).

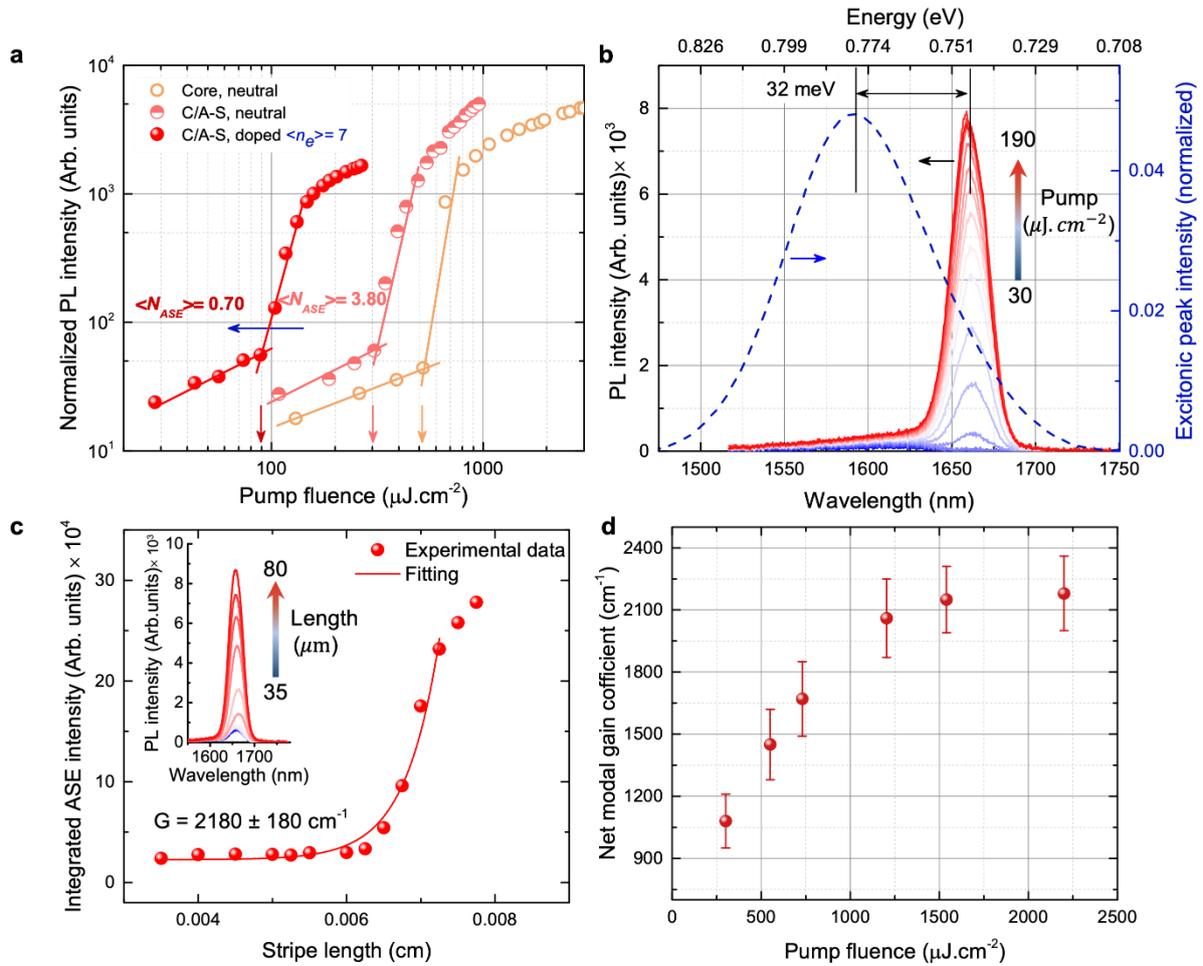

**Figure 3 | ASE and VSL characterizations of undoped- and doped -C/A-S CQDs. a,** Integrated intensity of ASE as a function of pump fluence for undoped-core, undoped-C/A-S and doped-C/A-S CQDs, with arrows indicating the ASE thresholds of ~ 500, ~ 300 and 90 µJ.cm$^{-2}$, respectively. The latter corresponds to $\langle N_{ASE} \rangle$= 0.70. **b,** Spectral analysis of ASE with respect to the lowest excitonic transition of undoped C/A-S CQDs (blue dashed line). The absorption peak is isolated from full spectrum (see Supplementary Fig.S6). The ASE peak is red-shifted by 32 meV compared to the first excitonic peak. **c,** Integrated intensity (ASE spectrum) of doped-C/A-S CQD thin film as a function of stripe length at a pump fluence of 2200 µJ.cm$^{-2}$. Inset shows the PL of the sample as function of stripe length. Based on analysis of VSL measurements, the net modal gain coefficient calculated as 2180± 180 cm$^{-1}$. **d,** The net modal gain coefficient obtained from VSL measurements varying the pump fluence. The maximum net modal gain is measured as 2180 cm$^{-1}$. Note, the error bars originated from the fitting.

d

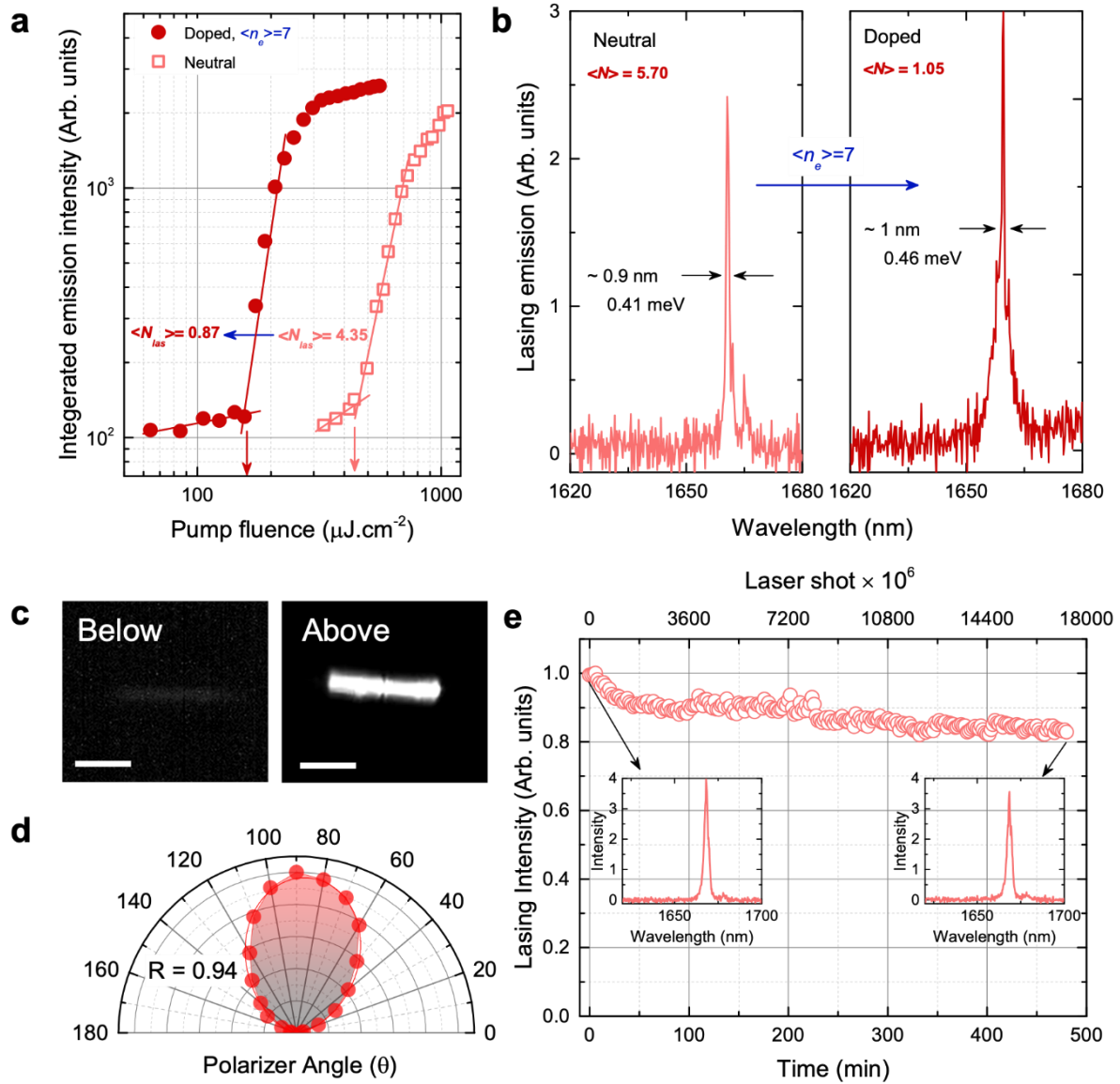

**Figure 4 | Sub-single-exciton infrared lasing demonstrated in doped-C/A-S CQDs device. a,** Integrated intensity of lasing peak of C/A-S CQDs (neutral & doped) coupled to a second-order DFB grating (out-plane emission) versus excitation intensity. Note, open squares represent neutral CQDs and filled circles are for doped CQDs. By applying permanent doping to the devices, $<n_e>$= 0 to 7, the lasing threshold decreases from ~430 to 160 μJ.cm$^{-2}$, for which the latter equals to $\langle N_{lasing} \rangle$= 0.87. **b,** Single mode lasing spectra for neutral- and doped- CQD/DFB devices at the corresponding fluences. Note, the exhibited spectra are just above the lasing threshold. **c,** Infrared images of the device below and above the lasing threshold. The scale bars, 0.5 mm. **d,** The polar plot represents the lasing emission intensity of the 1668 lasing mode as a function of the polarization angle (θ). Note, the solid line shows quadratic cosine function. "R" denotes the polarization ratio. **e,** Stability of lasing intensity over the course of 8 h of continuous excitation at 10 kHz. Insets show the lasing spectra at the corresponding time (arrows) of the exposure.

**Methods**

**Synthesis PbS core QDs**: The synthesis of PbS core QDs was described previously[2]. Briefly, 446 mg lead (II) oxide (PbO), 50 ml 1-octacedene (ODE) and 3.8 ml oleic acid (OA) were heated at 100°C under vacuum for 1h to form lead oleate. Under argon, and keeping the same temperature, a solution of 75µL hexamethyldisilathiane (HMS) dissolved in 3 mL ODE was quickly injected. After 6 minutes of reaction a second solution of 115µL HMS in 9 mL ODE was dropwise injected for 12 min. Afterwards, the solution was let cool down naturally. Once at room temperature it was washed three times with a mixture of acetone/ethanol, redispersing in toluene. The final sample was redispersed in degassed ODE and the concentration was adjusted to 70 mg/mL. This solution was stored at low temperature and under nitrogen.

**Synthesis of PbS@PbSSe core-shell QDs:** To grow a PbSSe shell on PbS core QDs, 240 mg PbO, 10 ml ODE and 1.1 ml OA were heated at 100°C under vacuum, for 1h, to form the lead precursor. After, the solution was changed to argon atmosphere, the temperature was set to 140°C and 2 mL of PbS-ODE solution were injected. When the temperature was recovered, the selenium precursor solution was injected (96 mg Se powder dissolved in 1.3 ml tetrabutylphosphine). The growth of the shell can be followed by absorption spectroscopy. When the thickness of PbSSe shell was the appropriated, the reaction was quenched cooling it down with a water bath. The QDs were washed three times with ethanol, redispersing in toluene. Finally, the concentration was adjusted to 30 mg/mL.

**QD film deposition.** QDs were deposited on the desired substrates by spin coting technique. First, a 30 mg/mL solution of QDs in toluene was spin casted for 20 s at a speed of 2,500 r.p.m. Then, a ligand exchange solution made of 7 mg/mL EMII in methanol with 0.03% MPA was poured to the surface of the formed solid film. The spin coating started after 30 s, at the same spin speed for 40 s. In order to clean the undesired organics remained from ligand exchange process, a few drops of methanol poured on the surface. The mentioned steps were repeated till reaching the desired film thickness. For doping QDs, the samples were subsequently transferred to ALD process with 30 cycles of trimethyl aluminum (TMA) and $H_2O$ at 60°C.

**TA characterizations.** The measurements were carried out employing a mode-lock Ti:sapphire femtosecond laser (45 fs) operating at 800 nm and a repetition rate of 1 kHz. The optical setup was designed as a typical pump-probe noncollinear configuration. The fundamental mode of the pump beam was directed into a half-wave plate and neutral density utilized to control the pump fluence. To generate the prob beam (ranging from 1200 to 1700 nm) an optical parametric amplifier used with approximately 1 mJ energy at 800 nm. In order to control the time delay between pump and probe beam, a precise motorized translation stage was used. The probe beam was focused on the sample as large as excitation beam area. The changes in reflection and transmission were simultaneously recorded.

**ASE and lasing measurements.** The samples were optically excited by a femtosecond (300 fs) Yt:YAG ORAGAMI laser (NKT photonics) operating at a wavelength of 1030 nm and a repetition rate of 10 kHz. A variable neutral density used for adjusting the pumping intensity. The laser focused through a cylindrical lens on the surface of the sample. The ASE spectra were collected perpendicular to the excitation axis by using a lens with a focal length of 50 mm having a diameter of 2 inches. Employing free-space optics, the collected ASE signal was coupled into a Kymera 328i spectrograph (Oxford instrument Andor) which equipped by an 1D-InGaAs camera via a lens having a focal length of 200 mm through a 100 µm slit.

Lasing spectra were recorded perpendicular to the surface of the devices ~ 20 cm away from the QD-DFB laser device. In doing so, a fiber coupled port of the spectrometer was used having a slit width of 10-20 µm to achieve high resolution. The integration time for all measurements set to 3 s. A long pass filter was located in front of the device for all measurements. The infrared picture and video of the QD/DFB laser were taken by a NIT-WiDy-SenS-320V-ST InGaAs camera with an attached SWIR lens.

**DFB grating fabrication.** Gratings were fabricated in a cleanroom by spin-coating PMMA (AR-P 662.04 Allresist GmbH) at 4000 rpm for 60 s on top of sapphire substrates (Ossila Ltd.) followed by a 2 min baking step at 150 °C. A conductive polymer layer (AR-PC 5090.02 Allresist GmbH) was then spin-coated on top at 2000 rpm and then baked at 90 °C. Samples were transferred to an electron beam lithography system (Crestec CABL 9000C) for patterning. After lithography, the conductive polymer layer was dissolved off in water for 60 s and the e-beam resist was developed for 1 min. Post development, spatially selective ALD deposition was performed to grow 50 nm TiO2 gratings between the PMMA structures. The residual PMMA was then removed using an oxygen plasma asher followed by acetone/isopropanol cleaning. DFB lasers were the fabricated by spin-coating 160 nm of CQDs on top gratings with the period of 860 and 870 nm.

**Acknowledgements**

G.K acknowledges support from H2020 ERC program grant number 101002306. The authors acknowledge support from the State Research Agency (AEI)/ PID2020-112591RB-I00/10.13039/501100011033 and PDC2021-120733-I00 funded by MCIN/AEI/10.13039/501100011033 by the European Union Next Generation EU/PRTR. Additionally, this project has received funding from the Spanish State Research Agency, through the CEX2019-000910 [MCIN/AEI/10.13039/501100011033], the CERCA Programme/Generalitat de Catalunya and Fundació Mir-Puig. This project has received funding from the European Union's Horizon 2020 research and innovation programme under the Marie Skłodowska-Curie grant agreement number 754558.

**Authors Contribution**

N.T. and G.K. conceived the concept of the study and G.K. supervised the study at all stages. N.T. and G.K. designed the experiments. N.T. fabricated, characterized the devices, and analyzed the data. M.D. synthesized PbS core and PbS/PbSSe core/shell QDs. G.W. fabricated the DFB gratings. A.O. and S.C. conducted the TA measurements. N.T. analyzed the TA data. N.T. and G.K. wrote the manuscript with input from co-authors.